\documentclass[twocolumn,showpacs]{revtex4}
\usepackage{epstopdf}
\usepackage{graphicx}% Include figure files
\usepackage{dcolumn}% Align table columns on decimal point
\usepackage{bm}% bold math

\def\gappeq{\mathrel{ \rlap{\raise.5ex\hbox{$>$}}
                      {\lower.5ex\hbox{$\sim$}}  } }
\def\lappeq{\mathrel{ \rlap{\raise.5ex\hbox{$<$}}
                      {\lower.5ex\hbox{$\sim$}}  } }
\begin{document}
\preprint{PRA}
\title{Collapse times of dipolar Bose-Einstein condensates}
\author{C. Ticknor$^1$, N.G. Parker$^2$, A. Melatos$^3$, S.L. Cornish$^4$, D.H.J. O'Dell$^2$ and A.M Martin$^3$}
\address{$^1$ ARC Centre of Excellence for Quantum-Atom Optics and Centre for Atom Optics and Ultrafast Spectroscopy, Swinburne University of Technology, Hawthorn, Victoria 3122, Australia. \\ $^{2}$ Department of Physics and Astronomy, McMaster University, Hamilton, Ontario, L8S 4M1, Canada. \\ $^{3}$ School of Physics, University of Melbourne, Parkville,
Victoria 3010, Australia. \\ $^4$ Department of Physics, Durham University, Durham, DH1 3LE, UK.}
\date{\today}
\begin{abstract}
We investigate the time taken for global collapse by a dipolar Bose-Einstein condensate. Two semi-analytical approaches and exact numerical integration of the mean-field dynamics are considered. The semi-analytical approaches are based on a Gaussian ansatz and a Thomas-Fermi solution for the shape of the condensate. The regimes of validity for these two approaches are determined, and their predictions for the collapse time revealed and compared with numerical simulations. The dipolar interactions introduce anisotropy into the collapse dynamics and predominantly lead to collapse in the plane perpendicular to the axis of polarization.  
\end{abstract}

 \maketitle

Long coherence times and a high degree of controllability make ultracold atomic gases suited to the study of non-equilibrium states of many-particle quantum systems.
One example is the collapse of a Bose-Einstein condensate (BEC) \cite{Sacket99,Gerton00,Donley01,Roberts01,Khaykovic02,Strecker02,Chin03,Cornish06}  by using a Feshbach resonance to change the $s$-wave scattering length $a_s$ from positive to negative.  Two limiting cases can be identified: global and local collapse depending, respectively, upon whether the (imaginary) healing length associated with $a_{s}$ is of the same order as, or much smaller than, the size of the BEC. The mechanism of global collapse is an instability of the monopole collective excitation mode which grows exponentially, causing the entire condensate to collapse in 3D \cite{sackett98}. Meanwhile, for local collapse it is high-lying phonon modes whose amplitudes grow fastest \cite{yurovsky}. Local collapse is expected when there is a sudden large change in $a_s$ within a large system \cite{Chin03}. The stability of a trapped BEC can be parametrized by $k_s=N a_s /a_\rho$ \cite{Roberts01,Ruprecht95,Gammal01,Parker07}, where $N$ is the number of atoms and $a_\rho$ is the radial harmonic oscillator length of the trap. The system collapses when the interactions are attractive ($k_s<0$) and $|k_s|$ exceeds a critical value $|k_{s}^c|$.   An important parameter defining collapse is the {\em collapse time}, $t_c$, which is the time taken for atomic three-body losses to become significant \cite{Donley01}.  Meanfield simulations including three-body loss have reproduced experimental results reasonably well \cite{savage}. 

 The $^{52}$Cr condensates made by the Stuttgart group  \cite{griesmaier} are the first to have large dipole-dipole interactions (in addition to the s-wave).  Dipolar interactions are long-range and partially attractive, and thus the properties of dipolar BECs (DBECs) are rather intriguing.  In recent experiments \cite{Koch07,Lahaye} the collapse of a DBEC was triggered by reducing the repulsive s-wave interactions, with the collapse proceeding anisotropically and on a global scale. Motivated by these experiments, we theoretically model the timescale for the global collapse of a DBEC.  This is achieved through mean-field simulations and two semi-analytic methods. %which offer ready-to-use convenience over full numerical simulations.  
 The first semi-analytic method is based on plasma physics treatments of the collapse of electrostatic \cite{Zakharov72,Robinson97} and electromagnetic \cite{Melatos07,Jenet07} wavepackets. We apply it to the case where the initial state is weakly interacting.  The second method is valid in the opposing interaction-dominated Thomas-Fermi (TF) regime; essentially we run the usual ballistic expansion equations \cite{Kagan96,Castin96} in reverse.

The macroscopic wave function describing a BEC $\psi({\bf r},t)$ satisfies
the Gross-Pitaevskii equation (GPE)
%A BEC can be described by a macroscopic wave function,
%$\psi({\bf r},t)$, via the Gross-Pitaevskii equation (GPE),
\begin{eqnarray}
i\hbar \frac{\partial \psi}{\partial t} &=&  \left[
-\frac{\hbar^2\nabla^2}{2m}  + V_T +g
\left|\psi\right|^2 + \Phi_{dd}
-i{\hbar K_3\over2}
|\psi|^4
\right]\psi, \nonumber\\\label{DGPE}
\end{eqnarray}
where $g=4 \pi \hbar^2 a_s/m$ and $m$ is the atomic mass. The confining harmonic potential, $V_T({\bf r})=m\omega_{\rho}^2\left(\rho^2+\gamma^2z^2\right)/2$, where $\rho^{2}=x^{2}+y^{2}$, is cylindrically symmetric, with radial (axial) frequency $\omega_\rho$
($\omega_z=\gamma \omega_\rho$).  The term $\Phi_{dd}({\bf r},t)=\int
d^3{\bf r^{\prime}}U_{dd}({\bf r}-{\bf r^{\prime}})n({\bf
r^{\prime}},t)$ accounts for the dipolar interactions $U_{dd}({\bf
r})=\mu_0 \mu^2(1-3\cos^2\theta)/(4 \pi |{\bf r}|^3)$,
where $\theta$ is the angle between the separation vector ${\bf r}-{\bf r^{\prime}}$ and the polarization direction, taken to be the $z$-direction.The strength of the dipolar interactions are characterized in terms of   $k_d=Na_{d}/a_{\rho}$, where $a_d=\mu_0\mu^2m/(12 \pi \hbar^2)$ is the length scale of the dipolar interactions. The last term in the GPE represents three-body loss where $K_3$ is the recombination rate. For the semi-analytic methods we assume that the DBEC collapses globally, maintaining a single-peaked density profile.  In the GPE simulations  some features of local collapse, discussed elsewhere  \cite{LongCollapse}, are observed.  This typically occurs after considerable global collapse and as such the dynamics remain in good agreement with global collapse predictions.  

Analogous collapse occurs in plasma
physics where electrostatic \cite{Zakharov72,Robinson97} and
electromagnetic \cite{Melatos07,Jenet07} wave packets in a turbulent
plasma undergo nonlinear self-focusing and Zakharov collapse.
Following approaches for plasma wavepacket dynamics
\cite{Robinson97}, and work by Lushnikov for DBECs \cite{Lushnikov}, the
equation of motion for the mean square radius of the DBEC, $\langle r^2\rangle=\langle \rho^2 +z^2\rangle$, 
is
\begin{eqnarray}
\partial_t^2 \langle r^2
\rangle
&=&\left[\frac{6E}{mN}-\frac{2E_{K}}{mN}-5\omega_{\rho}^2\langle r^2
\rangle
-  5\omega_{\rho}^2(\gamma^2-1)\langle z^2\rangle\right]  \nonumber \\
& &
%\partial_t^2 \langle r^2
%\rangle &=& \frac{1}{mN}\left[6E-2E_{KE}-5m\omega_{\rho}^2N\langle r^2
%\rangle \right. \nonumber \\
%&-& \left. 5m\omega_{\rho}^2N(\gamma^2-1)\langle z^2\rangle\right]
\end{eqnarray}
where $\langle r^2\rangle = \int r^2|\psi|^2 d^3{\bf r} $ and
$\langle z^2\rangle = \int z^2|\psi|^2 d^3{\bf r}$. We assume an
initial stable state with total and kinetic energy,
$E_0$ and $E_{K0}$, and, $\langle r^2 \rangle=\langle
r^2 \rangle_0$,  $\langle z^2 \rangle=\langle z^2 \rangle_0$ with
$d\langle r^2 \rangle_0/dt=d\langle z^2 \rangle_0/dt=0$.  Upon
changing the strength of the s-wave interactions from $a_{s0}$ to
$a_s$ the upper bound for the final value of $\langle r^2 \rangle$ is,
\begin{eqnarray}
\langle r^2
\rangle_f&\le&\langle r^2
\rangle_0+\frac{1}{mN}\left[6E^{\prime}-2E_{K0} - 5m\omega_{\rho}^2N\langle r^2
\rangle_0\right. \nonumber \\
 &-& \left. 5m\omega_{\rho}^2N(\gamma^2-1)\langle z^2\rangle_0\right]t^2,
\end{eqnarray}
where $E^{\prime}=E_0-E_{S0}(1-a_s/a_{s0})$. Rearranging, the  upper
bound for the collapse time is,
\begin{eqnarray}
t_c \le \left|\frac{\langle r^2
\rangle_f-\langle r^2
\rangle_0}{\frac{6E^{\prime}}{mN}-\frac{2E_{K0}}{mN}-5\omega_{\rho}^2\langle r^2
\rangle_0-5\omega_{\rho}^2(\gamma^2-1)\langle z^2\rangle_0}\right|^{\frac{1}{2}},% \nonumber \\
\label{upper_collapse}
\end{eqnarray}
where  collapse occurs when $\langle r^2\rangle_f=0$.

%  The total energy of the condensate ($E$) is evaluated using the GPE energy functional, which has four contributions:  kinetic ($E_K$), trap ($E_T$), s-wave interaction ($E_S$), and dipole ($E_D$) energies.

In the limit of weak interactions the condensate can be modeled
by a Gaussian ansatz (GA):
\begin{eqnarray}
\psi_{{\rm GA}}&=&\sqrt{{N}/{l_{\rho}^2l_z
\pi^{3/2}}}\exp\left(-\frac{\rho^2}{2l_{\rho}^2}-\frac{z^2}{2l_z^2}\right).
\label{P_Gauss}
\end{eqnarray}
The energy $E$ of this ansatz is evaluated using the GPE energy functional, and has four contributions:  kinetic ($E_K$), trap ($E_T$), s-wave ($E_S$), and dipole ($E_D$) \cite{yi01}. By minimising $E$ with respect to the radii $l_\rho$ and $l_z$  \cite{yi01,Pedri} the variational solutions for $E_0$, $\langle r^2
\rangle_f$ and $\langle r^2 \rangle_0$ are found. This enables the time it takes for the BEC to go from   $\langle r^2 \rangle_0$ to $\langle r^2
\rangle_f$  to be evaluated, via Eq.\ (\ref{upper_collapse}). For $\langle r^2
\rangle_f=0$  this defines the collapse time.
%and the energies given by the GA are well known \cite{Pedri} and can be used to infer a collapse time via Eq. (\ref{upper_collapse}).
%\begin{eqnarray}
%E_{K}&=&N\hbar\omega_{\rho}/2\left[1/l_{\rho}^2+1/2l_z^2\right]  \label{eq:E_K0} \\
%E_{T}&=&N\hbar \omega_{\rho}/2\left[l_{\rho}^2+\gamma^2 l_z^2/2\right] \\
%E_{S}&=&N\hbar \omega_{\rho}k_{s}/[\sqrt{2\pi}l_{\rho}^2l_z] \\
%E_{D}&=&N\hbar \omega_{\rho} k_{d}f\left(\kappa\right)/[\sqrt{2\pi}l_{\rho}^2l_z]. \label{eq:E_D0}
%\end{eqnarray}
%This ansatz can be used to infer a collapse time via Eq. \ref{upper_collapse}.

When the interactions become dominant it is then appropriate to use the TF approximation, wherein the zero-point kinetic energy is neglected. In this limit, the non-dipolar GPE under harmonic trapping is known to support an exact scaling solution given by \cite{Kagan96,Castin96},
\begin{eqnarray}
n({\bf r},t)&=&n_0(t)\left[1-\frac{\rho^2}{R_\rho^2(t)}-\frac{z^2}{R_z^2(t)}\right]  \label{density_TF}\\
{\bf v}({\bf r},t)&=&\frac{1}{2}\nabla\left[\alpha_\rho(t)\rho^2+\alpha_z(t)z^2\right] \label{vel_TF}
\end{eqnarray}
where $n_0(t)=15N/[8\pi R_\rho(t)^{2} R_z(t)]$. This class of solution remains exact even in
the presence of dipolar interactions \cite{ODell04,Eberlein05}. Substitution into the GPE yields equations of motion (\ref{lam_r}) and (\ref{lam_z}) for the radii  \cite{ODell04}. These describe the  two lowest-lying excitation modes, namely the axis-symmetric quadrupole and the monopole excitations.
%In the cylindrically symmetric case [$R_{x}(t)=R_{y}(t)=R_{\rho}(t)$] these  equations are constrained to model cylindrically-symmetric dynamics.
\begin{figure}
%\centering
\includegraphics[width=8.0cm]{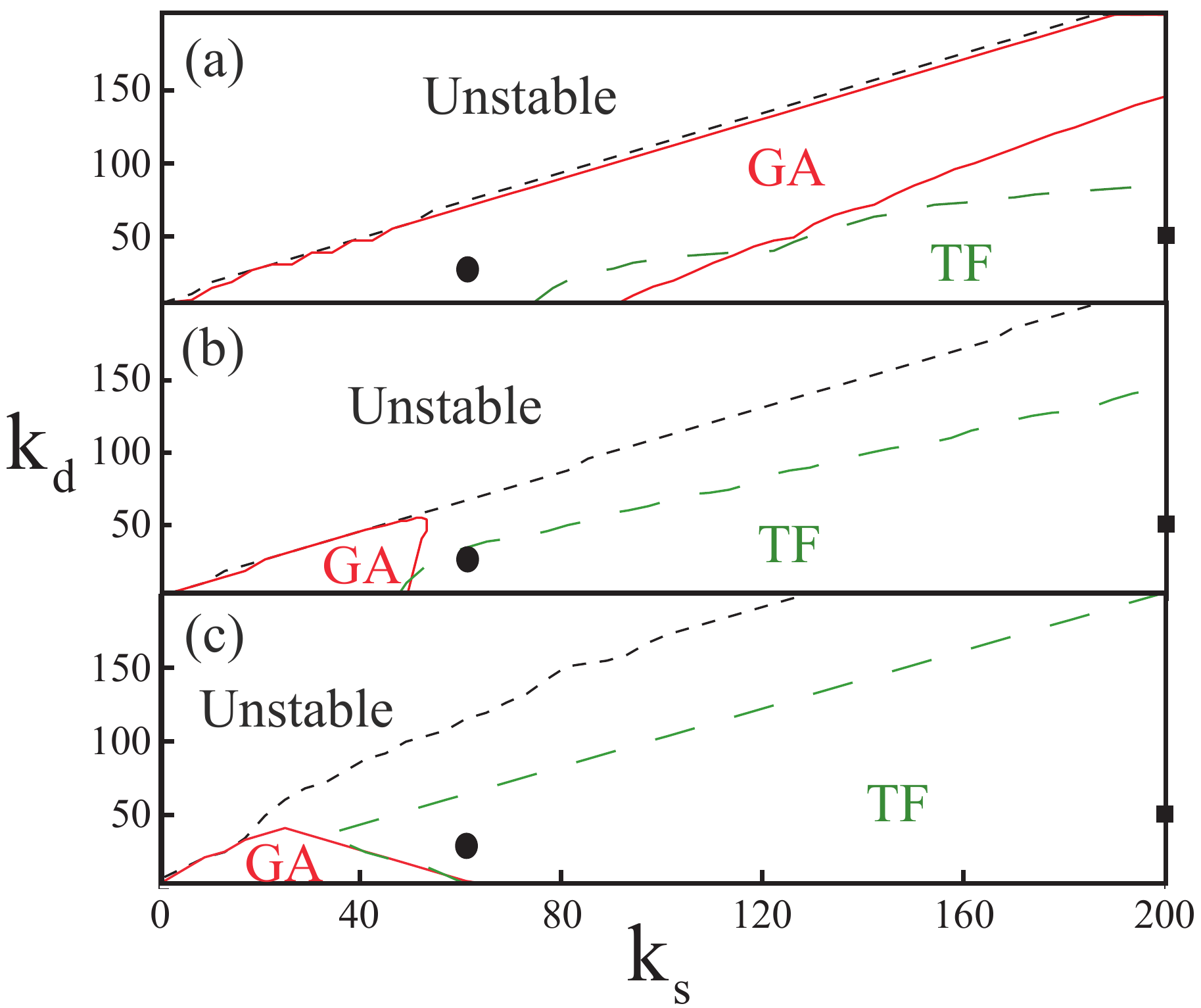}
\caption{The regimes of validity for the GA and TF ground states in the parameter space of $k_s$ and $k_d$.  Above the short dashed black curve  there are no stable ground states to the GPE.  The red solid (green long dashed) curve bounds the regions of validity of the GA (TF) solution, defined to be when their energy is within 5 $\%$ of that of the true GPE ground state.  Considered are the trap ratios of (a) $\gamma^2=0.1$, (b) $1$ and (c) $10$. In each of the figures the circles [squares] are the values for $k_{s0}$ and $k_d$ that are employed in Figs. 2(a-c) and Figs. 3(a-c) [Figs. 3(d-f)].  \label{region}}
 \vspace{-0.5cm}
\end{figure}

It is important to establish how well the TF and GA approximations reproduce the exact ground state in the parameter space of $k_s$ and $k_d$.  We define their regimes of validity to be when the energy of the GA or TF solution differs by less than $5\%$ from the energy of the  GPE solution.  Fig.\  \ref{region}  maps out these regimes for (a) $\gamma^2=0.1$, (b) $1$ and (c) $10$.
The short dashed black curve marks the threshold for collapse: above it there are no stable ground state solutions to the GPE.  Hence, for a given $k_d$ there is a unique $k_s^c$ below which the system is unstable to collapse. The region bounded by the solid red (long dashed green) curve marks the region of validity for the GA (TF) solutions.  In general we find that the GA gives a good approximation to the ground state for weak interactions and in regions close to the collapse threshold.   By contrast, the TF solutions are most accurate in the presence of strong s-wave interactions and for parameters well away from the collapse threshold.  Furthermore, trap geometry plays a key role in the validity of the solutions. In prolate, $\gamma<1$, traps the GA (TF) approximation has a large (small) region in which it is valid, while for oblate, $\gamma>1$, traps the opposite is true. This is because a prolate (oblate) dipolar BEC experiences a net attractive (repulsive) interaction due to the dominance of end-to-end (side-by-side) dipoles, and for such interactions the GA (TF) approximation works well. Note that close to the collapse threshold zero-point kinetic energy plays a significant role in stabilising the condensate and so the TF approximation does not provide a good description for ground states there.  However, the TF approach can still be employed to model the collapse {\em dynamics} providing that the interactions dominate over zero-point kinetic energy during the dynamics.  In practice this is achieved by beginning with an initial state that is well within the regime of stability (a TF initial state), and then suddenly switching to a point in the parameter space that is deep in the collapse regime, i.e. bypassing the threshold for collapse. As the density increases during collapse $E_{S}+E_{D}$ continues to dominate $E_{K}$.
%We find that although the GA works reasonably well for all the parameters considered, it  The GA region contains the origin, whereas the TF region contains large $k_s$ and small $k_d$. 
%It is important to note that the GA works reasonable well for all parameters 
%where there is a solution.  In contrast, the TF works poorly when the dipolar interactions are strong and significant distort the cloud.  This is why the
%TF region is always well below the black line.

We now study the dynamical collapse in the TF limit. We define the scaling parameters $\lambda_i(t)=R_i(t)/R_{i0}$, where $R_{i0}$ are the initial radii ($i=\rho, z$).  Then, under general time-dependent changes in $k_s(t)$ and $k_d(t)$, the time evolution of 
$\lambda_\rho(t)$ and $\lambda_z(t)$ is given by,
\begin{eqnarray}
\ddot{\lambda}_\rho=-\lambda_\rho&+&\frac{\alpha}{\lambda_\rho\lambda_z}\left[\frac{k_s(t)}{\lambda_\rho^2}\right. \nonumber \\
&-& \left.
k_d(t)\left(\frac{1}{\lambda_\rho^2}+3\kappa_0^2\frac{f(\kappa_0\lambda_\rho/\lambda_z)}{2(\kappa_0^2\lambda_\rho^2-\lambda_z^2)}\right)\right]  %\nonumber \\
  \label{lam_r}\\
\ddot{\lambda}_z=-\gamma^2 \lambda_z&+&\frac{\alpha\kappa_0^2}{\lambda_\rho^2}\left[\frac{k_s(t)}{\lambda_z^2} \right. \nonumber \\
&+& \left. k_d(t)\left(\frac{2}{\lambda_z^2}+\frac{3f(\kappa_0\lambda_\rho/\lambda_z)}{\kappa_0^2\lambda_\rho^2-\lambda_z^2}\right)\right], %\nonumber \\
 \label{lam_z}
\end{eqnarray}
where the unit of time is $1/\omega_\rho$,  $\alpha=15\kappa_0 a_\rho^5/R_{\rho0}^5$ and
%The function $f(\kappa)$ is
%\begin{eqnarray}
$f(\kappa)=\frac{1+2\kappa^2}{1-\kappa^2}-\frac{3\kappa^2{\rm atanh}\sqrt{1-\kappa^2}}{(1-\kappa^2)^{3/2}}$. 
The initial aspect ratio, $\kappa_0$, is evaluated from 
Eqs. (\ref{lam_r},\ref{lam_z}) for $\ddot{\lambda}_i=0$ \cite{yi01,ODell04}.

\begin{figure}
\centering
\includegraphics[width=8.0cm]{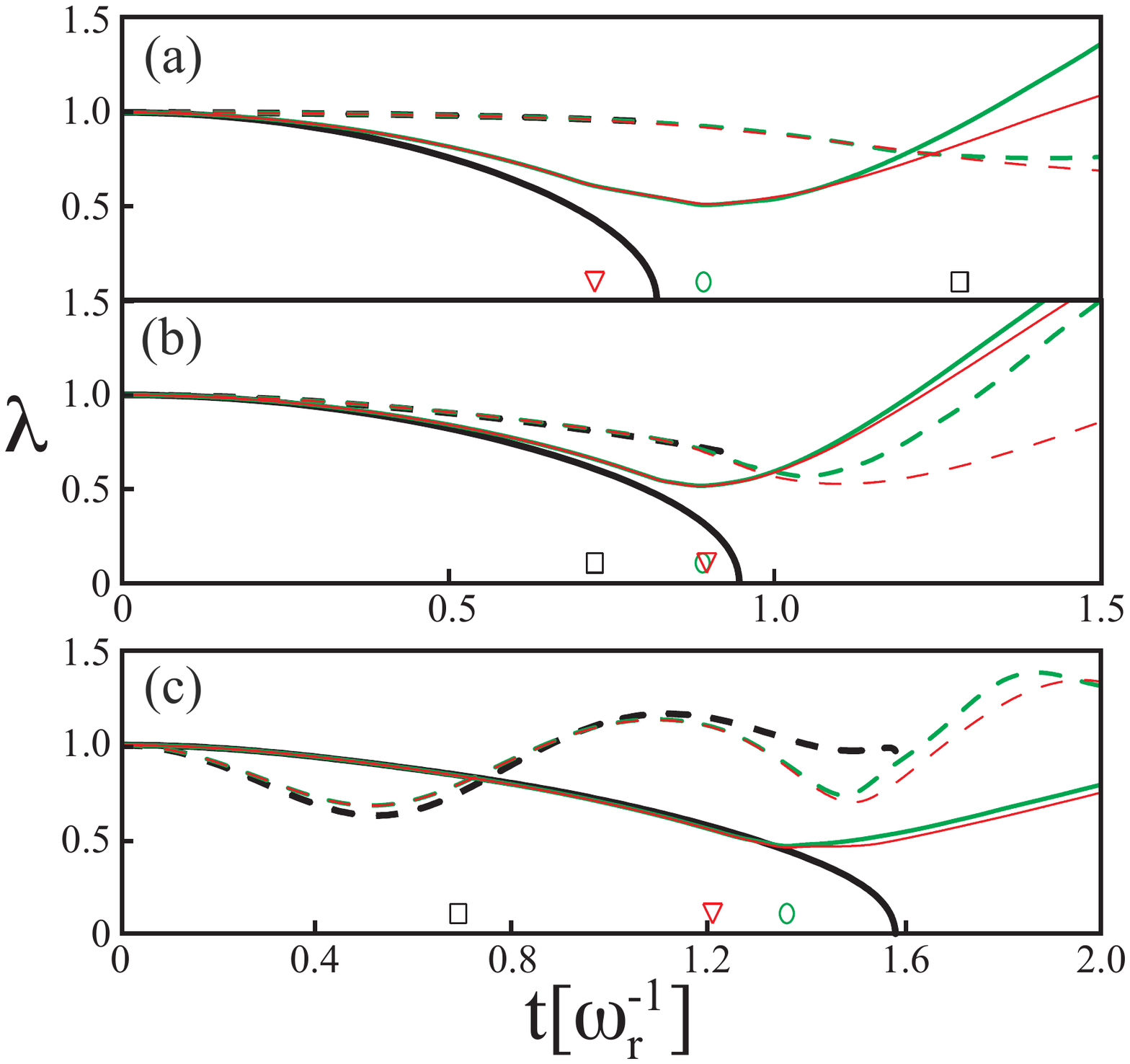} %, clip=true
\caption{$\lambda_\rho(t)$ and $\lambda_z(t)$ for $k_{s0}=60$, $k_s=0$ and $k_{d0}=k_d=30$, with (a) $\gamma^2=0.1$, (b) $1$, and (c)
$10$. The solid (dashed) curves correspond to $\lambda_\rho$ ($\lambda_z$);
thick black curves are solutions to Eqs. (\ref{lam_r},\ref{lam_z}) and the thin red (medium green) curves
are the full solutions to the GPE with (without) loss.
The red triangle indicates when the sudden onset of loss has begun, for $K_3=2\times10^{-28}$ cm$^6$/s \cite{Lahaye}. The black square is $t_c$ as derived from Eq. (\ref{upper_collapse}) using the GA and the green circle indicates when
$\dot\rho=0$ for the GPE solution in the absence of losses. For the GPE simulations we have defined
$\lambda_\rho^2(t)=\langle \rho^2(t)\rangle/\langle \rho^2(0)\rangle $ and $\lambda_z^2(t)=\langle z^2(t)\rangle/\langle z^2(0)\rangle $.   \label{lambda}}
\vspace{-0.5cm}
\end{figure}

%+++++++++FIG2+++++FIG2++++++++++++FIG2++++++++++++++++++++++++++++++++++++
Figure \ref{lambda} shows the evolution of $\lambda_\rho$ (solid curves) and $\lambda_z$ (dashed curves) as calculated from the TF equations (thick black curves) and numerical simulation of the GPE with (thin red) and without (medium green) three-body loss. Specifically,  the case where $k_s$ is switched from $k_{s0}=60$ to $0$ ($k_d=k_{d0}=30$) is considered, for (a) $\gamma^2=0.1$, (b) $1$ and (c) $10$. For these parameters the initial state of the BEC spans the regimes of validity of the TF and GA approaches. Comparing the GPE results to the TF analysis excellent agreement for the majority of the collapse is observed. Significant deviations only occur close to the point of collapse ($\lambda_\rho \rightarrow 0$) when the GPE collapse {\it bounces} and turns into an expansion of the system, consistent with the recent observation of a d-wave explosion following collapse \cite{Lahaye}.  Importantly, the collapse is highly anisotropic and occurs primarily in $R_\rho$, i.e. $R_\rho/R_z \rightarrow 0$ as $t \rightarrow t_c$, consistent with recent experimental observations \cite{Lahaye}.  The same behavior occurs in lower-hybrid collapse in turbulent plasmas \cite{Robinson97,Melatosa}. In the case of s-wave scattering, both $R_\rho$ and $R_z$ collapse at the same rate, with $R_\rho \propto (1-t/t_c)^{1/2}$, analogous to electrostatic Zakharov collapse \cite{Robinson97}. Furthermore, the fact that the TF parabolic scaling solutions give such a good description of the collapse indicates that, for the parameters considered, collapse is primarily a global effect and proceeds through a quadrupole collective mode.  
%The same behavior occurs in lower-hybrid collapse in turbulent plasmas \cite{Robinson97,Melatosa}. In the case of pure s-wave scattering, both $R_\rho$ and $R_z$ collapse at the same rate, with $R_\rho \propto (1-t/t_c)^{1/2}$, analogous to electrostatic Zakharov collapse \cite{Robinson97}. 
From the  GPE simulations a collapse time is defined in terms of a sudden onset of loss (red triangle) for $K_3>0$. Numerical simulations indicate that in the limit $K_3 \rightarrow 0$ this coincides with the time at which $\dot{\rho}=0$ (green circle).  %BUT IN FIGS 2A AND 2C THE GREEN ARROW DOES NOT COINCIDE WITH THE CIRCLE.... 
Hence, our results indicate that  GPE simulations without three-body losses can be used to infer an upper limit for the collapse time.  For pancake shaped geometries ($\gamma \gg 1$)  the TF and GPE solutions exhibit significant oscillations in  $\lambda_z$ before collapse, as seen in Fig. 2(c). 
%Additionally, for large attractive s-wave interactions collapse can occur first in $\lambda_z$ rather than  $\lambda_{\rho}$.

From Eq. (\ref{upper_collapse}), using the GA, an upper bound for the collapse can be calculated (black squares in Fig. 2). For $\gamma^2=0.1$ the upper bound, is consistent with GPE results. As we move away from the regime where the GA is valid, Figs. 2(b,c),  the upper bound for the collapse time  becomes inconsistent with the GPE results.    

\begin{figure*}
\centering
\includegraphics[width=16.0cm]{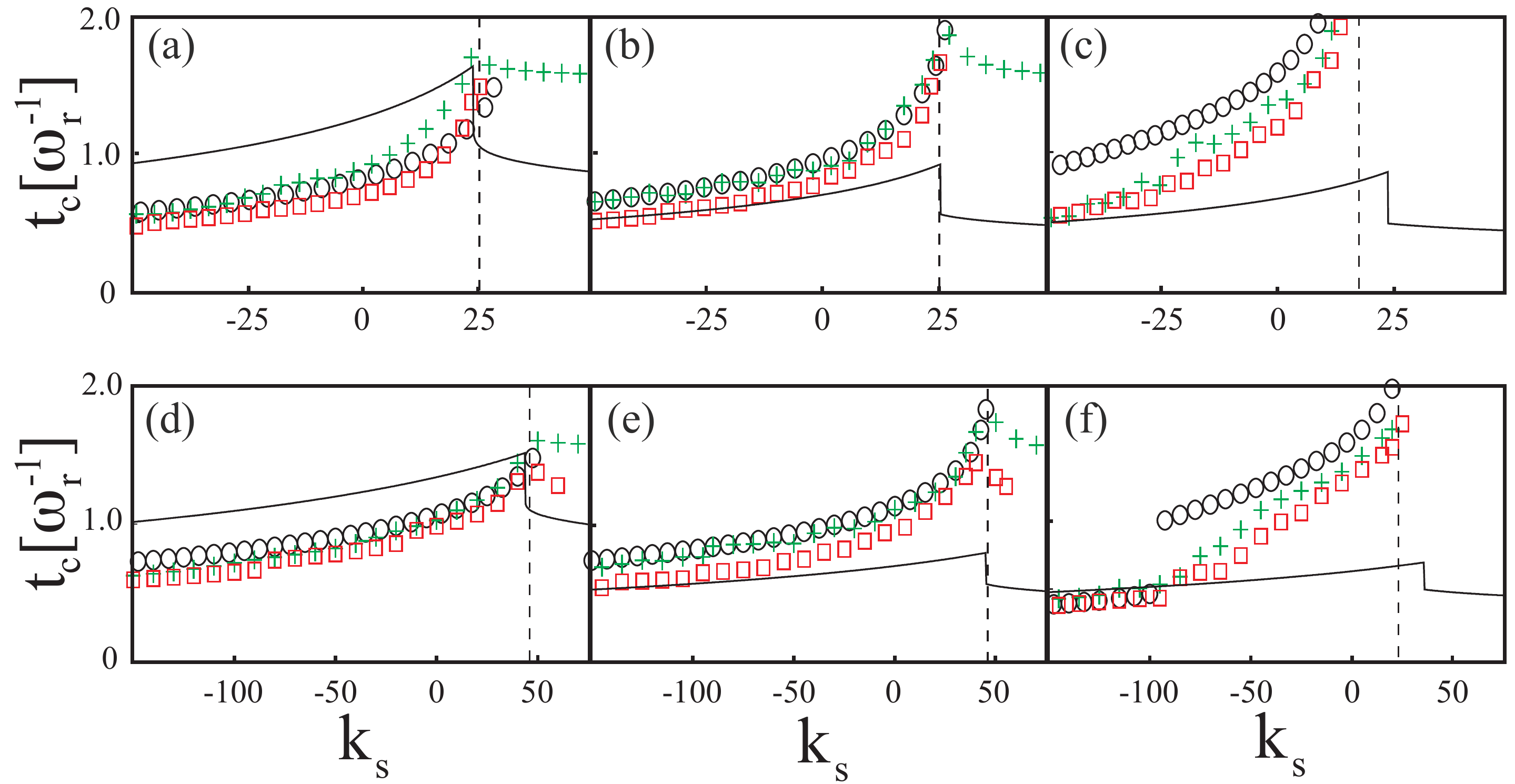}
\caption{Collapse time according to Eq. (\ref{upper_collapse}) (solid black curve),  Eqs. (\ref{lam_r},\ref{lam_z}) (black circles) and numerical integration of the GPE with (red squares) and without (green $+$, $\dot\rho$=0) 3-body loss. For numerical solutions of the GPE with (without) loss we have used $K_3=2\times10^{-28}$ cm$^6$/s \cite{Lahaye} ($K_3=0$).
(a-c) [(d-f)] $k_{s0}=60$ [200] and $k_{d0}=30$ [50], with (a) [(d)] $\gamma^2=0.1$, (b) [(e)] $1.0$ and (c) [(f)] $10.0$.  The vertical dashed lines correspond to $k_{s}^c$ as derived from static solutions to the  GPE.}
\vspace{-0.5cm}
\end{figure*}

Figure 3 presents the collapse times as $k_s$ is switched from $k_{s0}$ to $k_s$, for various geometries, as evaluated via Eq. (\ref{upper_collapse}) (solid black), Eqs. (\ref{lam_r},\ref{lam_z})  (black circles), and GPE simulations with (red squares) and without (green +) three-body losses. For $k_s < k_s^c$ [left of the vertical dashed lines], the final state solutions are collapsed states, with the time it takes to collapse increasing, as $k_s$ approaches $k^c_{s}$, from below.  In all of the regimes presented we find that the TF approximation provides a good estimate for the upper bound of the collapse time. In contrast, Eq. (\ref{upper_collapse}), is only consistent with the GPE simulations in the limit of very weak interactions.  Note the appearance of a sudden {\it step} in the TF and GPE collapse times in Fig.~3(f).  When the dipolar interactions dominate, the collapse is highly anisotropic and complete collapse occurs first in $R_{\rho}$. However, for large and attractive s-wave interactions, the collapse can become almost isotropic and for pancake-shaped systems complete collapse can occur first in $R_z$.  This step represents the transition between a complete collapse in $R_z$ (left of the step)  and $R_\rho$ (right of the step).

%{\bf Note, TF gives a very good estimate of when $\dot\rho=0$, and thus provides a reasonable upper limit time for collapse.}

%Ground state solutions are unstable to collapse for $k_{s0}<k^c_{s0}$ which corresponds to the region to the left vertical dashed line.  Under the sudden change in $k_s$, collapse approximately occurs in the same regime.  A general feature of is that as $k_s$ approaches $k^c_{s}$, from below, $t_c$ increases.  Note the appearance of a sudden `step' in the TF collapse times in Fig.~3(f).  When the dipolar interactions dominate, the collapse is highly anisotropic and complete collapse occurs first in $R_r$. However, for large and attractive s-wave interactions, the collapse can become almost isotropic and for pancake-shaped systems complete collapse can occur first in $R_z$.  This step represents the transition between a complete collapse in $R_z$ (left of the step)  and $R_\rho$ (right of the step).

%Varying agreement is observed between the GA and TF approaches, and the full numerical simulations.  This can be explained by considering the regimes of validity of the TF and GA methods. 

%TF does well in predicting when the system's critical $k_s$ of collapse.

In summary we have presented two simplified models of the collapse time of a DBEC, and compared them with exact numerical integration of dipolar GPE.  When the atomic interactions are weak or attractive a GA for the initial DBEC can be assumed with an upper bound for  the collapse time derived through a highly simplified equation of motion for the radius [Eq.~(4)].  In the opposing regime where interactions dominate, TF equations of motion for the radii provide a convenient and approachable method to derive the time for global collapse.   The validity of these two regimes is determined by the strength of the interactions and  the aspect ratio of the trap. When dipolar interactions dominate, the collapse occurs in the plane perpendicular to the axis of polarization.  The excellent agreement, for the parameters considered, between the TF dynamics and the full numerical simulations indicates that the collapse primarily occurs in a global manner and proceeds through a quadrupolar collective motion, consistent with the recent experimental observations  \cite{Koch07,Lahaye}.  Finally, for oblate geometries, we observe two prominent deviations from this general behavior. Firstly, significant oscillations in $\lambda_z$ can occur during collapse and secondly, for strong attractive s-wave interactions the collapse predominantly occurs in $z$ rather than $\rho$.  

This work was funded by the ARC (CT,AM,AMM), Canadian Commonwealth Scholarship Program (NGP), EPSRC and Royal Society (SLC) and NSERC (DHJOD).

\end{document}